\documentclass[useAMS,usenatbib,usegraphicx]{mn2e}
\usepackage{graphicx}
\usepackage{amssymb}
\usepackage{textcomp}
\usepackage{array}
\usepackage{multirow}
\usepackage[T1]{fontenc}

\newcommand{\aap}{A\&A}
\newcommand{\aj}{AJ}
\newcommand{\apj}{ApJ}
\newcommand{\apjl}{ApJ}
\newcommand{\apjs}{ApJS}

\newcommand{\araa}{ARA\&A}
\newcommand{\mnras}{MNRAS}

\newcommand{\pasj}{PASJ}
\newcommand{\pasp}{PASP}

\newcommand{\na}{New Astronomy}

\newcommand{\shortitle}{IFS study of the ionized gas in Stephan's Quintet}

\usepackage[ps2pdf,bookmarksopen,breaklinks=true]{hyperref}

\hypersetup{citecolor={black},linkcolor={black},urlcolor={black}}  
\hypersetup{
  colorlinks=true,             
  bookmarksnumbered=true,      
  bookmarksopen=true,         
  pdfpagemode=UseThumbs,     
  pdfdisplaydoctitle,          
  pdftitle={\shortitle}
}

\DeclareRobustCommand{\ion}[2]{%
\relax\ifmmode
\ifx\testbx\f@series
{\mathbf{#1\,\mathsc{#2}}}\else
{\mathrm{#1\,\mathsc{#2}}}\fi
\else\textup{#1\,{\mdseries\textsc{#2}}}%
\fi}

\newcommand{\hh}{\ion{H}{ii}~}
\newcommand{\hei}{\ion{He}{i}}

\newcommand{\nii}{[\ion{N}{ii}]}
\newcommand{\oi}{[\ion{O}{i}]}
\newcommand{\oii}{[\ion{O}{ii}]}
\newcommand{\oiii}{[\ion{O}{iii}]}
\newcommand{\sii}{[\ion{S}{ii}]}

\title[\shortitle]{A study of the ionized gas in Stephan's Quintet from
  integral field spectroscopy observations\thanks{Based on observations collected 
  at the Centro Astron\'omico Hispano Alem\'an (CAHA) at Calar Alto, operated
  jointly by the Max-Planck Institut f\"ur Astronomie and the Instituto de
  Astrof\'isica de Andaluc\'ia (CSIC).}}
\author[Rodr\'{i}guez-Baras et al.]{M.~Rodr\'{i}guez-Baras$^1$\thanks{E-mail: marina.rodriguez@uam.es}, 
  F.~F.~Rosales-Ortega$^{2,1}$, A.~I.~D\'{i}az$^1$, S.~F.~S\'{a}nchez$^{3,4}$,
  \and A.~Pasquali$^5$\vspace{0.3cm}\\
$^1$ Departamento de F\'{i}sica Te\'orica, Universidad Aut\'onoma de Madrid, 28049 Madrid, Spain.\\
$^2$ Instituto Nacional de Astrof{\'i}sica, {\'O}ptica y Electr{\'o}nica, Luis E. Erro 1, 72840 Tonantzintla, Puebla, Mexico\\
$^3$ Instituto de Astronom\'\i a,Universidad Nacional Auton\'oma de Mexico, A.P. 70-264, 04510, M\'exico,D.F.\\
$^4$ Instituto de Astrof\'{i}sica de Andaluc\'{i}a (CSIC), Camino Bajo de Hu\'{e}tor s/n, Aptdo. 3004, E18080-Granada, Spain\\
$^5$ Astronomisches Rechen-Institut, Zentrum f\"{u}r Astronomie der Universit\"{a}t Heidelberg, M\"{o}nchhofstrasse 12-14, 69120 Heidelberg, Germany\\
}

\begin{document}

\date{\today}

\pagerange{\pageref{firstpage}--\pageref{lastpage}} \pubyear{}

\maketitle

\label{firstpage}

\begin{abstract}
The Stephan's Quintet (SQ) is a famous interacting compact group of galaxies in an important stage of dynamical evolution, but surprisingly very few spectroscopic studies are found in the literature. We present optical integral field spectroscopy (IFS) observations of the SQ from the PPAK IFS Nearby Galaxies Survey (PINGS), that provide a powerful way of studying with spatial resolution the physical characteristics of the ionized gas within the group. The nature of the gas emission is analysed using 2D maps of continuum-subtracted, pure emission-line intensities, stacked spectra, diagnostic diagrams, and photoionization model predictions. In the case of NGC\,7319, we study the galaxy-wide high-velocity outflow emission by comparing the emission properties with theoretical shock and AGN models. We conclude that the excitation mechanism of the gas in this galaxy is a mixture of AGN photoionization and shocks with a photoionizing precursor. The emission spectra from the large scale shock front in the interacting pair NGC\,7318A/B is analysed, confirming the presence of several kinematic components. Comparison with predictions from theoretical shock models suggests that the gas emission is consistent with shocks without a photoionizing precursor, low pre-shock density, and velocities in the range of $\sim200-400$ km~s$^{-1}$. The gas oxygen abundance for NGC 7318B is determined using an updated O3N2 calibration. Although NGC\,7317 shows no significant gas emission, an ionizing cluster is found southwest of this galaxy, probably the result of tidal interaction. As a by-product, we analyse the gas emission of the foreground galaxy NGC\,7320.
\end{abstract}



\section{Introduction}
\label{intro}

\begin{figure*}
\centering
\includegraphics[bb=0 -10 245 215,scale=1.3]{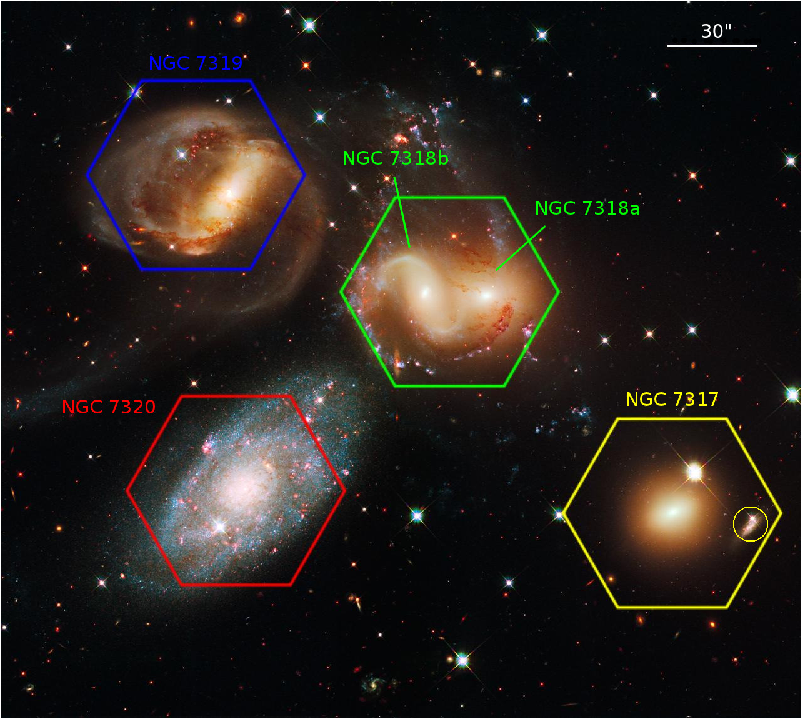}
\caption{
  HST image of Stephan's Quintet, in standard astronomical orientation (North up, East left). Overlaid there are shown the four hexagonal
  pointings of this project: blue for NGC\,7319, red for NGC\,7320, green for
  NGC\,7318A/B and yellow for NGC\,7317. The yellow circle in NGC\,7317 pointing indicates the region discussed in Sec.\ref{discussion7317}. Credit: NASA; ESA and the Hubble SM4
  ERO Team. A colour version of this figure can be found in the online version of the article.}
\label{pointings}
\end{figure*}

Compact galaxy groups at low redshift are ideal laboratories for studying the effects of extreme galaxy interactions, which are believed to be very important in driving galaxy evolution at high redshift. One of the best examples is the Stephan's Quintet (hereafter SQ), a
strongly interacting compact group of six galaxies (see Fig. \ref{pointings})
that has been the subject of many studies since it was discovered in 1877 by
M.E. Stephan. It is an ideal candidate for detailed analysis, as it is bright
and is in a rare but important stage of dynamical evolution that allows us to
observe manifestations of both past and present interaction events.

The SQ includes a core of three galaxies (NGC\,7317, NGC\,7318a and NGC\,7319)
with essentially zero velocity dispersion that have previously undergone
several episodes of interactions. These past events are probably due to the
dynamical harassment of a fourth galaxy with accordant redshift, NGC\,7320c, called the ``old intruder'' in the dynamical scenario described by
\citet{1997ApJ...485L..69M}. Both NGC\,7317 and NGC\,7318a are ellipticals, while
NGC\,7319 and NGC\,7320c (which is not shown in Fig. 1) show spiral  morphology but have lost almost all
their interstellar medium (ISM) along successive processes of interaction
within the group \citep{2001AJ....122.2993S}. Now the gas-rich spiral NGC\,7318b,
the ``new intruder'', is apparently entering the group for the first time at
high velocity ($\approx$1000 km s$^{-1}$), colliding with the intergalactic
medium of the group \citep{2001AJ....122.2993S,2003ApJ...595..665X}. The sixth object,
NGC\,7320, presents a highly discordant redshift \citep{1961ApJ...134..244B}
and is known to be a foreground dwarf spiral galaxy. Many prominent elements
of the SQ morphology are consequence of the interaction history of the group,
like the unrelaxed stellar halo comprising 30\% of the optical light \citep{1998A&A...334..473M} or the two tidal tails of different ages pointing to the old
intruder NGC\,7320c \citep{2001AJ....122.2993S}. But the most striking feature of
the group is the galaxy-scale shock generated by the ongoing collision between
NGC\,7318b and the intra-group gas, which is the tidally stripped ISM from
NGC\,7319 and NGC\,7320c during previous interactions. Along the last decade
this shock ridge has been detected and studied in a wide range of wavelengths: from X-ray \citep{2003A&A...401..173T,2005A&A...444..697T} to radio through its synchrotron emission \citep{2001AJ....122.2993S,2002AJ....123.2417W}. Excitation diagnostics from optical \citep{2003ApJ...595..665X} and mid-IR
\citep{2010ApJ...710..248C} emission lines also confirm the presence of shocked gas,
and Spitzer mid-IR observations reveal that the mid-IR spectrum in the shock
structure is dominated by the rotational line emission of molecular hydrogen,
H$_{2}$ \citep{2006ApJ...639L..51A,2010ApJ...710..248C}. Recent integral field spectroscopy \citep{2012A&A...539A.127I} of the ionized gas in this shock has revealed the presence of several kinematic components.

Despite being the most studied compact group, there are many questions about
the Quintet that remain unsolved. Relatively few spectroscopic studies have
been carried out, some of them dealing with the general properties of the group as its
dynamical state \citep{1998A&A...334..473M} and others focusing on particular
aspects \citep{1996AJ....111..140A,2004A&A...426..471L,2010ApJ...710..248C,2012ApJ...748..102T}, but there is still much lacking information about the complex
phenomena that define the past and future of these galaxies. Integral field
spectroscopy (IFS) may imply a significant step further in this field, as it
allows a complete mapping of the physical properties of the relevant regions
of the group. Hitherto only \citet{2012A&A...539A.127I} has presented
IFS observations of the SQ, but they focused specifically on the shock region
between NGC\,7318b and the debris field. In this paper we present for the
first time 2D spectroscopic observations with a wide field of view that
completely include five of the six galaxies related to the SQ, as well as some
of the intergalactic regions. This provides a powerful way of studying with
spatial resolution the physical characteristics of the ionized gas both in the
galaxies and in the intra-group material generated by the
interactions. Based on a proven methodology which includes the use of
complete maps of pure emission-line intensities, the anaylysis on extracted stacked
spectra from specific regions of interest, and comparison with theoretical model
predictions, we confirm and extend previous results about the nature of the
outflows in the Seyfert 2 galaxy NGC\,7319, the properties of the shock region
associated with NGC\,7318b, the discovery of current star formation in
NGC\,7317, and as a by-product, the internal structure of the foreground
galaxy NGC\,7320. This work is focused on the gas component of the SQ, but these observations also provide very interesting information about the stellar populations, whose study will be performed elsewhere. The paper is organized as follows: in Sec.\ref{Observations}
we describe the observations and data reduction. Sec.\ref{discussion} presents
the analysis performed on the IFS data and the main results obtained. The main
conclusions are summarized in Sec.\ref{conclusions}.


\section{Observations and data reduction}
\label{Observations}

The SQ observations are part of the PPAK IFS Nearby Galaxy Survey (PINGS,
\citealt{2010MNRAS.405..735R}), a project designed to construct 2D spectroscopic
mosaics of a representative sample of nearby spiral galaxies. PINGS
observations were carried out at the 3.5m telescope of the Centro
Astron\'{o}mico Hispano Alem\'{a}n (CAHA) at Calar Alto with the Postdam Multi
Aperture Spectrograph (PMAS; \citealt{2005PASP..117..620R}) in the PPAK mode \citep{2004AN....325..151V,2006NewAR..50..355K,2006PASP..118..129K}, i.e. a retrofitted bare fibre
bundle IFU which expands the FoV of PMAS to a hexagonal area with a footprint
of 74x65 arcsec$^{2}$. The PPAK unit contains 331 densely packed optical
fibres to sample an astronomical object at 2.7 arcsec per fibre, with a
filling factor of 65\% due to gaps in between the fibres. The sky background
is sampled by 36 additional fibres distributed in six mini-IFU bundles of six
fibres each. Additionally, 15 fibres can be illuminated directly by internal
lamps to calibrate the instrument. During the observations the V300 grating
was used, covering a wavelength between 3700 and 7100 \AA\ with a resolution
of 10 \AA\ FWHM, corresponding to 600 km s$^{-1}$ for \oiii~$\lambda$5007 and
460 km s$^{-1}$ for H$\alpha$. The exposure times per field for PINGS non-dithered frames, including also the SQ observations, were 3 x 600 s. More details about PINGS observations are available in \citealt{2010MNRAS.405..735R}. In the case of Stephan's Quintet, four
individual pointings were observed on the photometric night of 2008 August 10, three of which were
centred at the bright bulges of NGC\,7317, NGC\,7319 and NGC\,7320, while the
last pointing was centred to cover NGC\,7318a and
NGC\,7318b. The seeing was always smaller than 1.3 arcsec, never comparable to the fibre size. The FoV of the PPAK instrument is shown in Fig. \ref{pointings} as
hexagons overlaid on the HST image, labelled according to the galaxy or galaxies
they cover. The whole data set comprises 1324 fibre spectra.

\begin{figure*}
\centering
\includegraphics[bb=42 187 510 800]{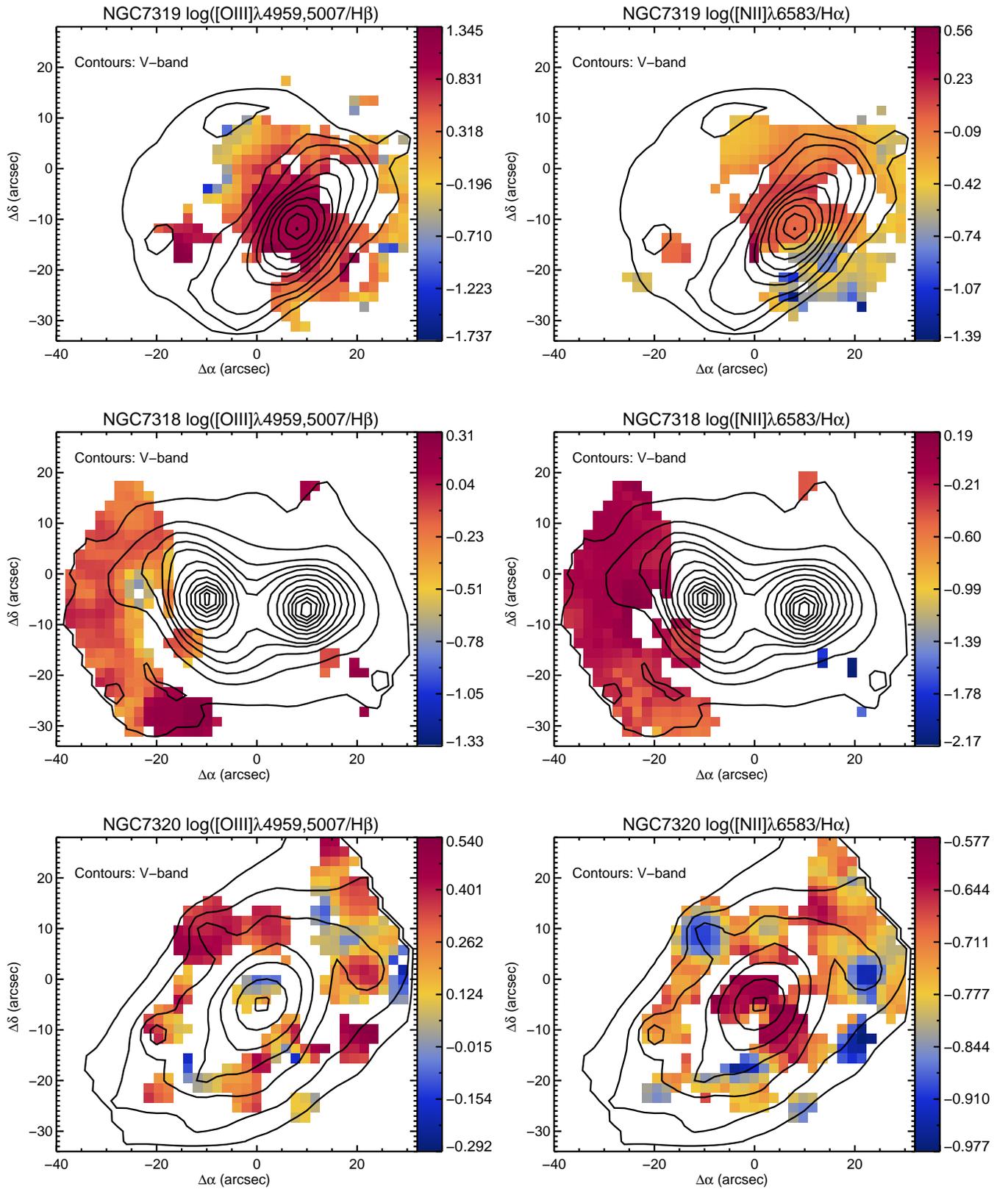}
\caption{
  Examples of emission-line ratio maps derived from the IFS cubes of
  NGC\,7319, NGC\,7318A/B and NGC\,7320. Left-column:
  \oiii~$\lambda$5007/H$\beta$ maps. Right column:
  \nii~$\lambda$6584/H$\alpha$. Maps are displayed with a standard orientation
  north-east positive. A colour version of this figure can be found in the online version of the article.}
\label{maps}
\end{figure*}

The pre-reduction processing was performed using standard IRAF \footnote{IRAF
  is distributed by the National Optical Astronomy Observatories, which are
  operated by the Association of Universities for Research in Astronomy, Inc.,
  under cooperative agreement with the National Science Foundation.} packages,
while the main reduction was performed using the R3D software for fibre-fed
and IFS data \citep{2006AN....327..850S} in combination with the E3D visualization
software \citep{2004AN....325..167S}. After that, the fibre-based IFS data was spatially resampled to a
datacube with a regular grid of 2 arcsec/spaxel, adopting a flux-conserving, natural-neighbour, non-linear interpolation, following the scheme described in \citet{2012A&A...538A...8S}, developed for the CALIFA survey. Astrometric corrections were applied to the final
data cube, and foreground objects (stars) were eliminated from the FoV of the
pointings. Then,
following the procedure outlined in \citet{2010MNRAS.405..735R} and
\citet{2011MNRAS.410..313S}, the {\sc FIT3D} software \citep{2007A&A...465..207S}
was used to fit the underlying stellar population of each spaxel spectrum by a
linear combination of single stellar population (SSP) templates from the MILES
library \citep{2010MNRAS.404.1639V} covering a wide range of ages (0.09, 1.00 and
17.78 Gyr) and metallicities (Z $\sim0.0004, 0.03$). Once the stellar
continuum is subtracted, a pure-emission spectrum is obtained. Based on this
residual spectrum, the intensity of the emission lines is calculated in each
spaxel by fitting one or multiple Gaussian profiles to the most prominent
emission lines in the considered wavelength range, including: H$\alpha$,
H$\beta$, H$\gamma$, \oii~$\lambda$3727, \oiii~$\lambda\lambda$4959,5007, \nii~$\lambda\lambda$6548,6583 and \sii~$\lambda\lambda$6717,6731, and also for the most
prominent sky residuals present in the spectrum, i.e. \oi~$\lambda$5577, and
\ion{Na}{i}~$\lambda$5893 lines (see \citealt{2011MNRAS.410..313S}, for more details).


\section{2D spectroscopic analysis of the Stephan's Quintet}
\label{discussion}

\subsection{Emission line ratio maps.}
\label{discussion1maps}

The measurement of the line intensities for every spaxel provides images of
the spatial distribution of all the strong emission line species
included in the wide wavelength range of our observations. We applied a flux
threshold in order to avoid regions dominated by low S/N data. For this
purpose, we produced masks considering two criteria. Firstly, we empirically
determined for each pointing the value of the H$\alpha$ flux below which the
spaxels are clearly not associated with the galactic structure and can be associated with the
background, and eliminated these spaxels. We did the same for the 
H$\beta$ map. Secondly, we discarded those regions that were not consistent
with the expected theoretical value of the H$\alpha$/H$\beta$ ratio (2.87
following \citealt{2006agna.book.....O}, within the systematic errors in the
measurement of Balmer lines). With this masking process we reproduced the galactic structure as seen in the HST image (see Fig.\ref{pointings}) and we made sure that we were considering
regions that are physically meaningful.

\begin{figure*}
\centering
\includegraphics[bb=37 186 850 520, scale=0.60]{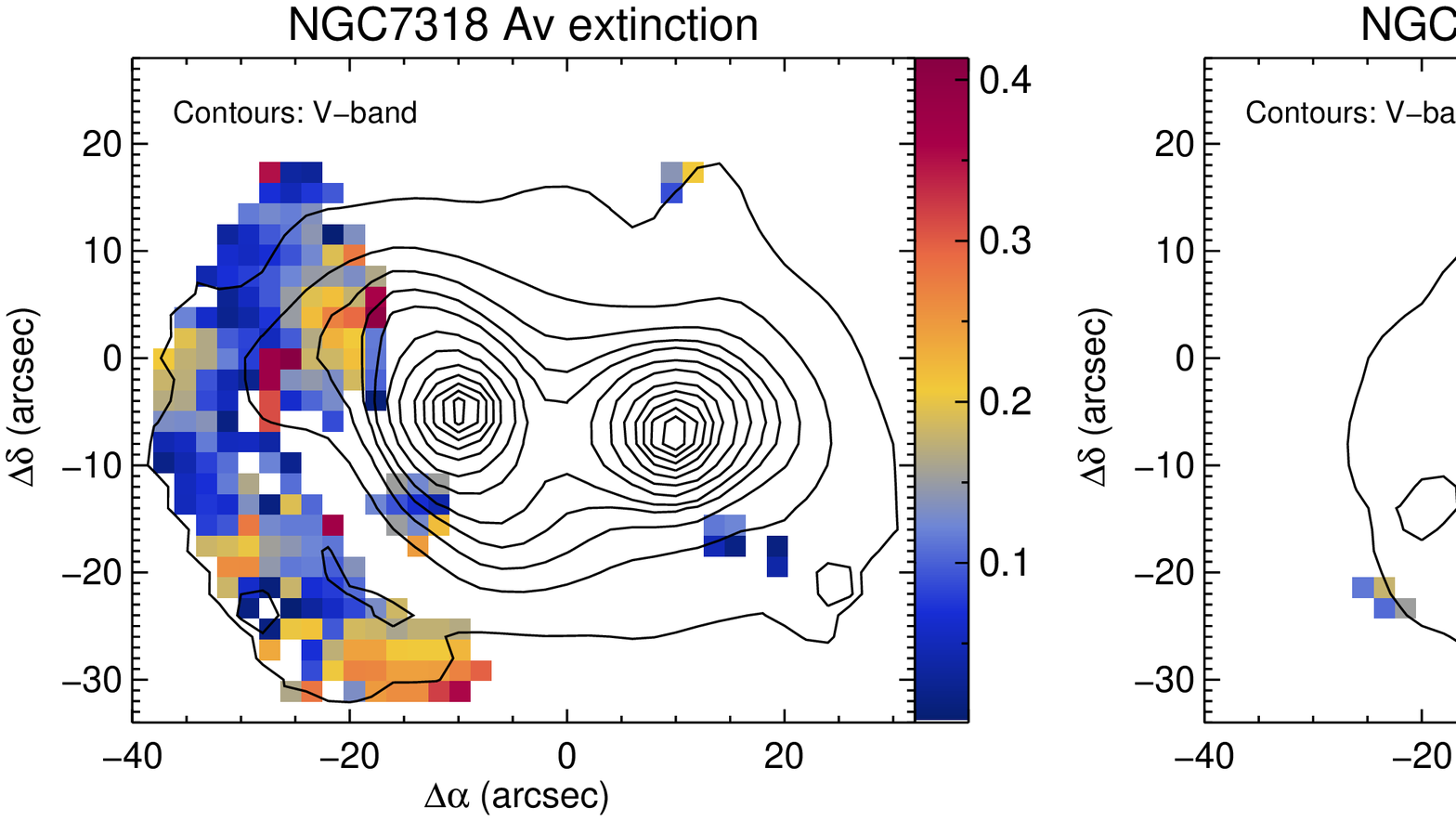}
\caption{Left: A$_{V}$ extinction map of NGC\,7318A/B, obtained from the Balmer decrement (H$\alpha$/H$\beta$) using the reddening function of \citet{1989ApJ...345..245C}. Right: A$_{V}$ extinction map of NGC\,7319, obtained as indicated for NGC\,7318A/B. A colour version of this figure can be found in the online version of the article.}
\label{extinctionfigure}
\end{figure*}

\begin{figure*}
\centering
\includegraphics[bb=61 188 850 1050, scale=0.6]{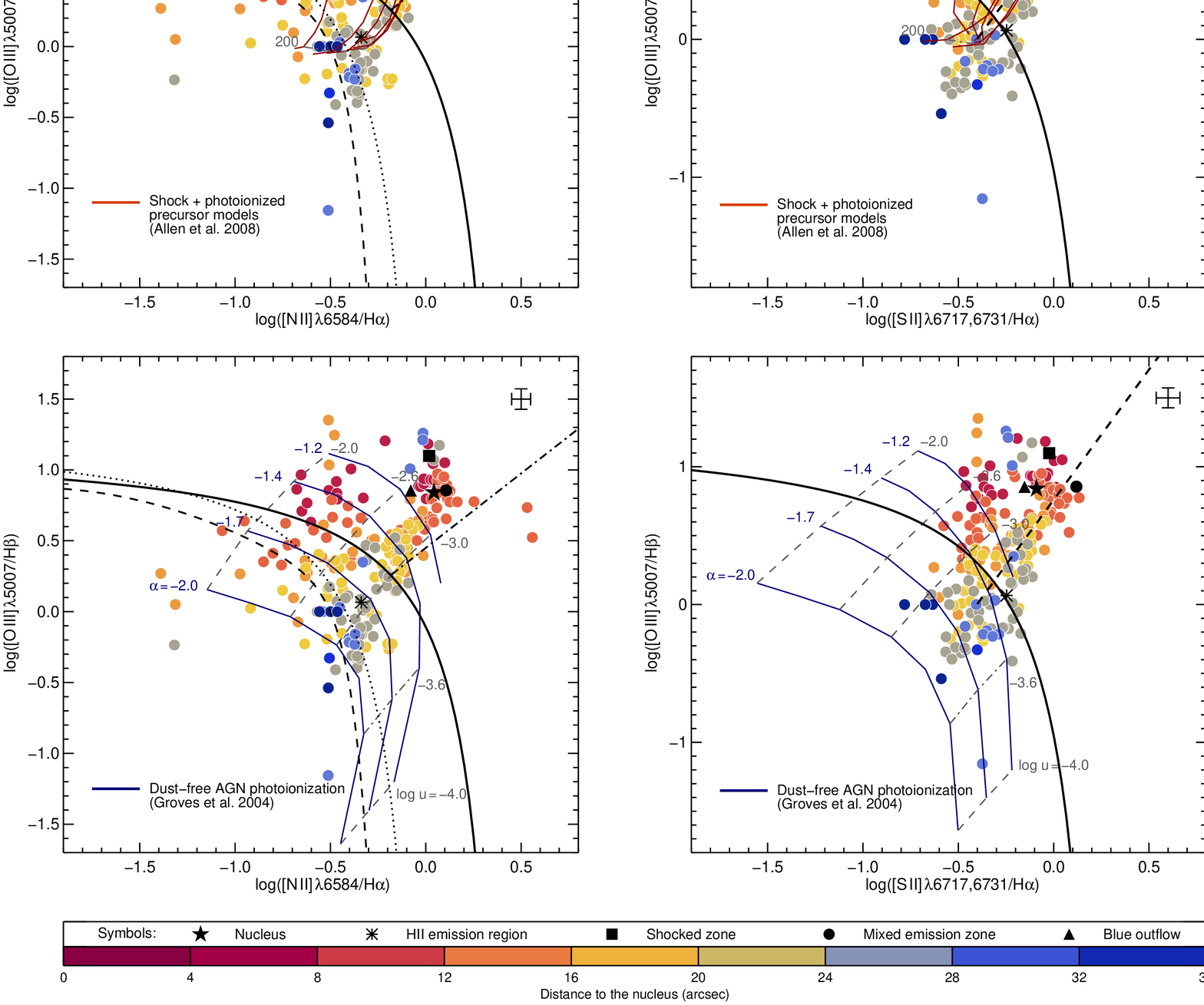}
\caption{Line ratio diagnostic diagrams, showing line ratios for independent
  spaxels in NGC\,7319. Line ratios in the key regions of interest (see
  Fig. \ref{7319total}) are plotted with black symbols, as indicated. Overplotted
  as black lines are empirically and theoretically derived separations between
  LINERs/Seyfert and \hh regions. Overplotted as coloured lines are line
  ratios predicted for photoionization of gas by a fast shock with precursor
  from \citet{2008ApJS..178...20A} and by AGN without dust from \citet{2004ApJS..153...75G},
  as described in the text. A colour version of this figure can be found in the online version of the article.}
\label{7319bpts}
\end{figure*}

This procedure was applied to all the pointings except NGC\,7317, which
showed no gas emission except for a region outside the galaxy that will be
discussed in Sec. \ref{discussion7317}. For the other three pointings we used
the masked images of the emission lines to represent maps of key line ratios sensitive
to ionization state, focusing on ratios of lines that are close enough in
wavelength not to be strongly affected by dust attenuation. The species which
are presented as the ratio with respect to H$\alpha$ were masked with the
corresponding H$\alpha$ mask, while species presented as the ratio with respect to
H$\beta$ were masked with the corresponding H$\beta$ mask. Some examples of
the final maps are shown in Fig. \ref{maps}, together with V-band contours
of the corresponding galaxy or galaxies obtained with {\sc PINGSoft}
\citep{2011NewA...16..220R} after applying a V filter to the data, for comparison between the shape of the galaxy in the continuum light and the gas distribution. The maps show the galactic structures and highlight a wide range of phenomena according to each context, like the
AGN with shock emission in NGC\,7319, the shock front in NGC\,7318A/B or the
\hh regions distribution in NGC\,7320, that will be further discussed in the
following sections.  

The Av extinction maps of NGC\,7318 and NGC\,7319, included in Fig. \ref{extinctionfigure}, are derived from the Balmer decrement of each spaxel spectrum, according to the reddening function of \citet{1989ApJ...345..245C}, assuming $R \equiv A_V /E(B - V ) = 3.1$. Theoretical value for the intrinsic line ratio H$\alpha$/H$\beta$ was taken from \citet{2006agna.book.....O}, assuming case B recombination (optically thick in all the Lyman lines), an electron density of $n_{e}$=100 cm$^{-3}$ and an electron temperature $T_{e}$=10$^{4}$K. In the case of NGC\,7319, we detect an extinction of $\sim0.5-1$ mag in the nucleus and in the major part of the galaxy, consistent with dust produced by intense star formation in these regions. We notice that the higher extinction is found at the edges of the emission region, probably tracing dust accumulated by the gas expansion, although we should take into account that the extinction calculated in those regions may be conditioned both by the weak H$\beta$ emission and the limitations in the fitting of the stellar population models that might not correctly reproduce the H$\beta$ profiles in absorption. In the case of NGC\,7318 extinction map, we find a relatively low extinction with average values between $0.15-0.25$ mag, in a very inhomogeneous distribution, consistent with the mixing and turbulence expected in the shocked region.


\begin{figure*}
\centering
\includegraphics[bb=30 50 842 500, scale=0.65]{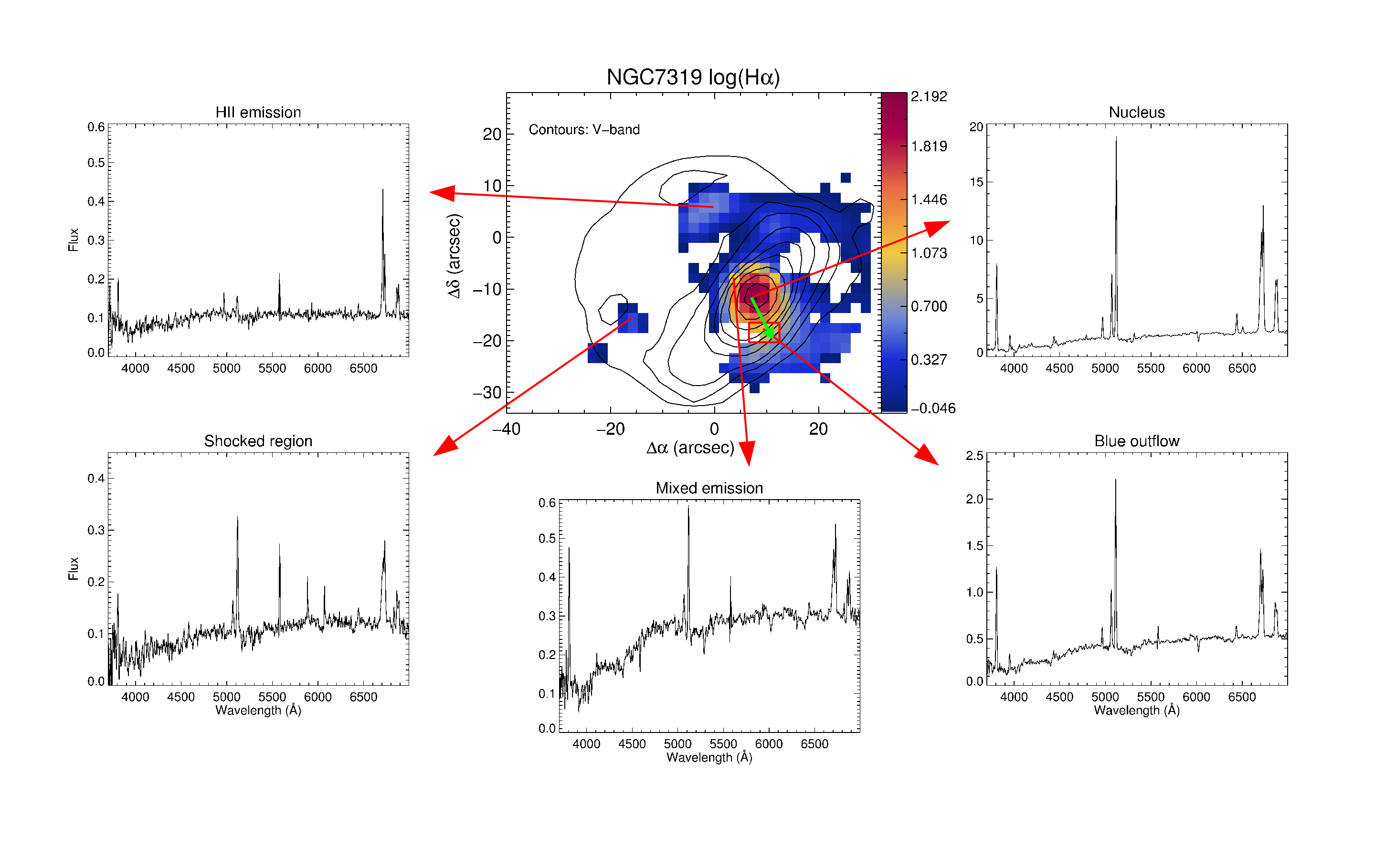}
\caption{Stacked spectra, obtained with {\sc PINGSoft}, from several key
  regions of NGC\,7319. The red box shows the spaxels where the blueshift in
  H$\alpha$ is resolved. The green arrow shows the
  trajectory along which the evolution of the H$\alpha$ line profile is
  studied in Fig. \ref{7319evol}. Fluxes are expressed in units of 10$^{-16}$ erg s$^{-1}$ cm$^{-2}$. (The prominent feature observed in some spectra at $\sim 5500$ \AA\ is the \ion{O}{i}~$\lambda$5577 sky emission line, the most visible at Calar Alto.) A colour version of this figure can be found in the online version of the article.}
\label{7319total}
\end{figure*}

\subsection{The Seyfert 2 galaxy NGC\,7319}
\label{discussion7319}

This galaxy has played a main role in all the interaction episodes within
the group, losing most of its gas as a consequence. These interactions may be
responsible for the activity of its nucleus, classified as Seyfert 2
\citep{2006A&A...455..773V}. 
A powerful way to probe the nature of the dominant ionizing source in galaxies
are the emission-line diagnostic diagrams (introduced by \citealt*{1981PASP...93....5B}; hereafter BPT). They work by exploring the location of certain
line ratios, involving several strong emission lines with a dependence on the
ionization degree and, to a lesser extent, on temperature or
abundance. Through the application of different classification criteria
\citep{2001ApJ...556..121K,2006MNRAS.372..961K,2003MNRAS.346.1055K,2006MNRAS.371..972S,2007MNRAS.382.1415S} diagnostic diagrams allow the separation of galaxies
into those dominated by ongoing star formation and the ones dominated by
non-stellar processes, and with sufficient information can further split the latter 
into Seyfert AGN and LINERs. Some of the diagrams also contain a transition
region, where the classification method indicates a blend of star formation
and AGN activity.

The spatial resolution of the line emission maps obtained from the IFS
observations of the SQ has allowed us to elaborate diagnostic diagrams with the particular
positions of every spaxel, instead of studying the integrated emission of each
galaxy as a whole. The classification of each spaxel according to its position
in the diagnostic diagrams provides information to study the distribution of the
different ionizing sources across the galaxy. In the case of
NGC\,7319, the diagnostic diagrams highlight the coexistence of several
processes, like the AGN emission from the NGC\,7319
nucleus and the shocked gas emission as a consequence of the outflow from the nucleus (see Fig.\ref{7319bpts}). The integrated spectrum of
the nuclear region was extracted using {\sc PINGSoft}, by co-adding the spectra of the
spaxels corresponding to that zone in the data cube. In order to obtain the
pure-emission spectrum we applied the {\sc FIT3D} software to the extracted
spectrum, to fit and subtract the underlying stellar continuum, as explained
in Sec. \ref{Observations}. In Fig. \ref{7319total} we plot examples of
stacked spectra from several key regions of the galaxy, corresponding to
different physical scenarios that will be discussed along this section. 
The spectrum of the nucleus, included in Fig. \ref{7319total}, confirms the
classification of this galaxy as Seyfert 2.

\begin{figure}
\centering
\includegraphics[bb=67 219 668 1010, scale=0.38]{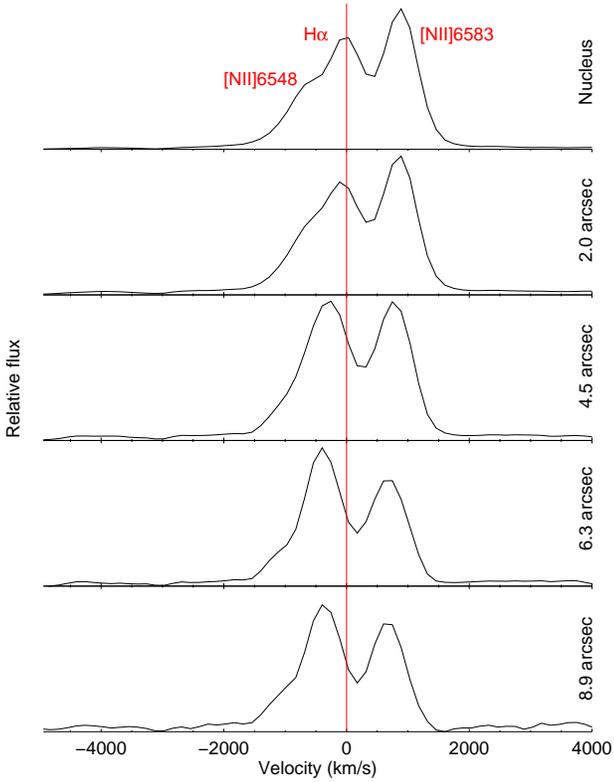}
\caption{Variations of the H$\alpha$ profile along the green arrow drawn in
  Fig. \ref{7319total}, from the galactic nucleus to the outflow region with
  velocities of the order of 500 km s$^{-1}$. Each spectrum corresponds to the extraction of the emission of one spaxel. Distances to the nucleus are indicated on the right.}
\label{7319evol}
\end{figure}

\citet{1996AJ....111..140A} (hereafter, A96) reported the presence of a high-velocity
outflow, that was later the objective of a first attempt of study with IFS
techniques by \citet{2003MSAIS...3..226B} and \citet{2008ASPC..396...61D}. Using
long-slit optical spectroscopic observations, A96 found that the gas in the
south-west region relative to the nucleus is blueshifted from the systemic
velocity by 300 km s$^{-1}$ in average, and 500 km s$^{-1}$ at maximum. After
discarding galactic rotation and tidal effects as possible causes, it was
finally concluded that this blueshift is generated by an outflow that may have
a red counterpart in the north-east region, although this could not be
observed with their data. The PPAK V300 PINGS observations spectral resolution (see Sec.\ref{Observations}) does not allow to study the average velocity variations described by A96. This is
not a surprising fact, as the observations were not intended to perform a
kinematic analysis, but to cover a wide range in wavelength. Nevertheless we
do detect some spaxels with blueshifted values of the order of 500 km s$^{-1}$
in H$\alpha$, located in a region of $6" \times 4"$ in the south-west at
about 5'' from the nucleus that is signaled with a red box in
Fig. \ref{7319total}. The variation of the H$\alpha$ line profile from the
nucleus to this south-west region along the green arrow included in
the central map of Fig. \ref{7319total} is represented in
Fig. \ref{7319evol}. This is in very good agreement with the maximum blueshift
reported by A96. Observations with higher spectral resolution would be
necessary to construct a complete radial velocity map and to study the
possible presence of a red outflow located north-east from the nucleus. Comparison with observations in other wavelength ranges would also be advisable. A simple comparison between our maps and those obtained by \citet{2004MNRAS.353.1117X} reveals similarities between the emission structures observed in visible and radio ranges, leading to concordant conclusions about the presence of jets and shock fronts around the nucleus of the galaxy.

\begin{figure}
\centering
\includegraphics[bb=27 177 523 600, scale=0.43]{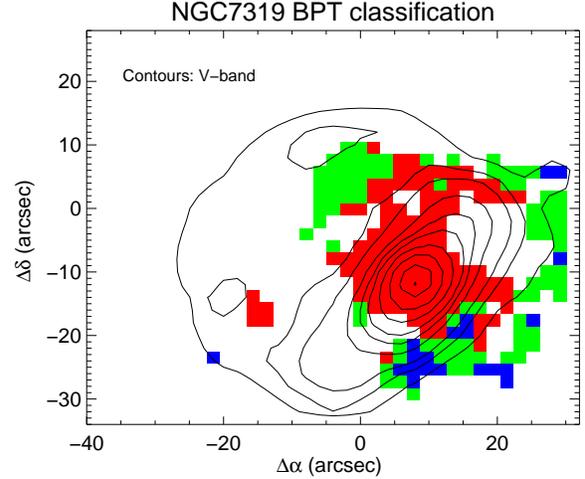}
\caption{NGC\,7319 spatial map with the spaxels classified according to their
  position in the \oiii~$\lambda$5007/H$\beta$ vs \nii~$\lambda$6584/H$\alpha$
  diagram. Spaxels classified as AGN are displayed in red, spaxels classified
  as starburst are displayed in blue, and spaxels classified as composite are
  displayed in green. A colour version of this figure can be found in the online version of the article.}
\label{7319bptmap}
\end{figure}

However, the outflow effects that we are not able to fully detect in the velocity field
are instead much more visible in the study of the measured emission-line
ratios. We observe features in the maps that would correspond to shocks caused
by the outflowing material coming from the nucleus. In the optical range,
shock-ionized regions can be distinguished from photo-ionized gas (\hh
regions) through their strong lines from low excitation species, such as
\oi~$\lambda$6300, \oii~$\lambda$3727, \nii~$\lambda\lambda$6548,83 and
\sii~$\lambda\lambda$6717,31 relative to H$\beta$. Temperature sensitive ratios
(e.g. \oiii~$\lambda$5007/H$\beta$) are also good indicators of shock activity
since shocked gas can readily attain high temperatures. These two facts
together place these shocked regions in the LINER/Seyfert location in
diagnostic diagrams. In general, LINER-like emission line ratios are observed
in the case of shocks, and if the shock is travelling quickly enough, a
photoionized precursor can increase the ionization state of the gas even
further, leading to Seyfert-like line ratios \citep{2008ApJS..178...20A,1996ApJS..102..161D}. Fig. \ref{7319bptmap} shows a galaxy map classifying the
spaxels according to their position in the \oiii~$\lambda$5007/H$\beta$ vs
\nii~$\lambda$6584/H$\alpha$ BPT diagram (see Fig. \ref{7319bpts}, left
panels). Spaxels placed above and to the right from the parametrization line
by \citet{2001ApJ...556..121K}, and therefore classified as AGN (both Seyfert and
LINER), are coloured in red. Spaxels below the parametrization line by
\citet{2006MNRAS.371..972S}, classified as starburst, are coloured in
blue. Spaxels located between these two lines, classified as composite, are coloured in green. This composite classification has no physical meaning for individual \hh regions, but has a high interest in cases that are supposed to respond to a mixture of the two possible emission
mechanisms (often associated with shocks). The
central region appears coloured in red, as expected, and the green zones,
i.e. the ones representing composite spaxels, delimit a circle surrounding the
AGN-like emission region. Only a few spaxels are coloured in blue, in
accordance with the absence of gas associated with pure starforming regions. This
scenario corresponds precisely to the hypothesis of shocks generated by the
outflowing material from the nucleus. Using the information provided by the
diagnostic diagrams and applying shock and AGN theoretical models \citep{2008ApJS..178...20A,2004ApJS..153...75G} we attempted to determine the main
characteristics of these shocks. 

According to the treatment of interstellar shocks, assuming a frame of reference
that is co-moving with a driving shock into the surrounding medium, in the
thin shock front the kinetic energy of the gas is converted into thermal
energy through elastic collisions \citep{1993ARA&A..31..373D,2005pcim.book.....T}. In the
postshock relaxation layer, this energy will cause ionization processes. As
shocks are supersonic with respect to the gas upstream and subsonic relative
to the gas behind, shock waves cannot travel ahead of the shock, but the shock
may be preceded by a radiative precursor, affecting the pre-shock gas. In
high velocity shocks (v$_{s}$ $>$ 50 km s$^{-1}$), this effect will lead to
photoionization. The recombination photons, i.e. extreme UV and soft X-ray
photons which are generated by the cooling of hot gas behind the shock front,
may travel upstream and pre-ionize the pre-shock gas. 

The emission spectrum of radiative shocks depends upon the shock velocity and
the physical conditions in the pre-shock material. The observed line
intensities and widths in the outflow of NGC\,7319 are consistent with the
spectra of radiative shocks with velocity v$_{s}$ $>$ $\sim100$ km s$^{-1}$
\citep{1993ARA&A..31..373D}. Typically, the shock velocity (v$_{s}$) can be
diagnosed using H$\alpha$, \oi~$\lambda$6300, \oii~$\lambda$3727 and
\oiii~$\lambda$5007 based on the emission-line velocity dispersion $\sigma$
and/or the FWHM. Nevertheless, as it was previously discussed, we do not have
enough spectral resolution to determine the velocity dispersion of the gas or
to resolve any line splitting in the data. Therefore, since the scenario
proposed by A96 seems consistent with our observations, we use the velocity
range obtained by A96 (300-500 km s$^{-1}$) as a first approach to the shock
velocity in order to estimate the parameters of the theoretical models that
will be compared with our data.

Another important physical quantity that has an effect on the final emission
spectrum of radiative shocks is the density of the pre-shock material. The
average electron density may be measured from the ratio of excitation lines such as
\sii~$\lambda\lambda$6717/6731 (e.g. \citealt{2006agna.book.....O}). The \sii emission
arises downstream, in the relaxation layer where the temperature is
$\sim10^{4}$K, so the \sii derived average electron density ($n_{SII}$)
would correspond to these regions. In the fast shock limit there is a density
jump by a factor of 4 between the pre-shock material and the region immediately
behind the thin shock front \citep{1993ARA&A..31..373D}, i.e $n_{1}=4n_{0}$, and in
the post-shock region (with $n_{2}=n_{SII}$ and $T_{2}\sim10^{4}$ K), the gas
cooling is approximately isobaric \citep{2005pcim.book.....T}. Therefore we can calculate
an approximation of the pre-shock density assuming isobaric conditions
downstream the shock front (i.e. $n_{1}T_{1}=n_{2}T_{2}$), considering
$T_{2}\sim8000$ K which is typical of \sii\, emission, and using the observed
$n_{2}$ derived from the measurement of the \sii~$\lambda\lambda$6717/6731 ratio in
the shock region, the preliminary velocity range of 300-500 km s$^{-1}$ from
A96, and

\begin{equation}
T_{1} \sim 1.4 \cdot 10^5 \left(\frac{v_{s}}{100 km s^{-1}}\right)^2,
\end{equation}

\noindent
which is the temperature immediately behind the shock front for a fully
ionized gas (assuming an ideal gas and 0.1 helium fraction, \citealt{2003adu..book.....D}) Through this procedure we derive for our velocities range a
pre-shock density range of $\sim 1-10$ cm$^{-3}$.

\begin{figure*}
\centering
\includegraphics[bb=0 10 842 524, scale=0.6]{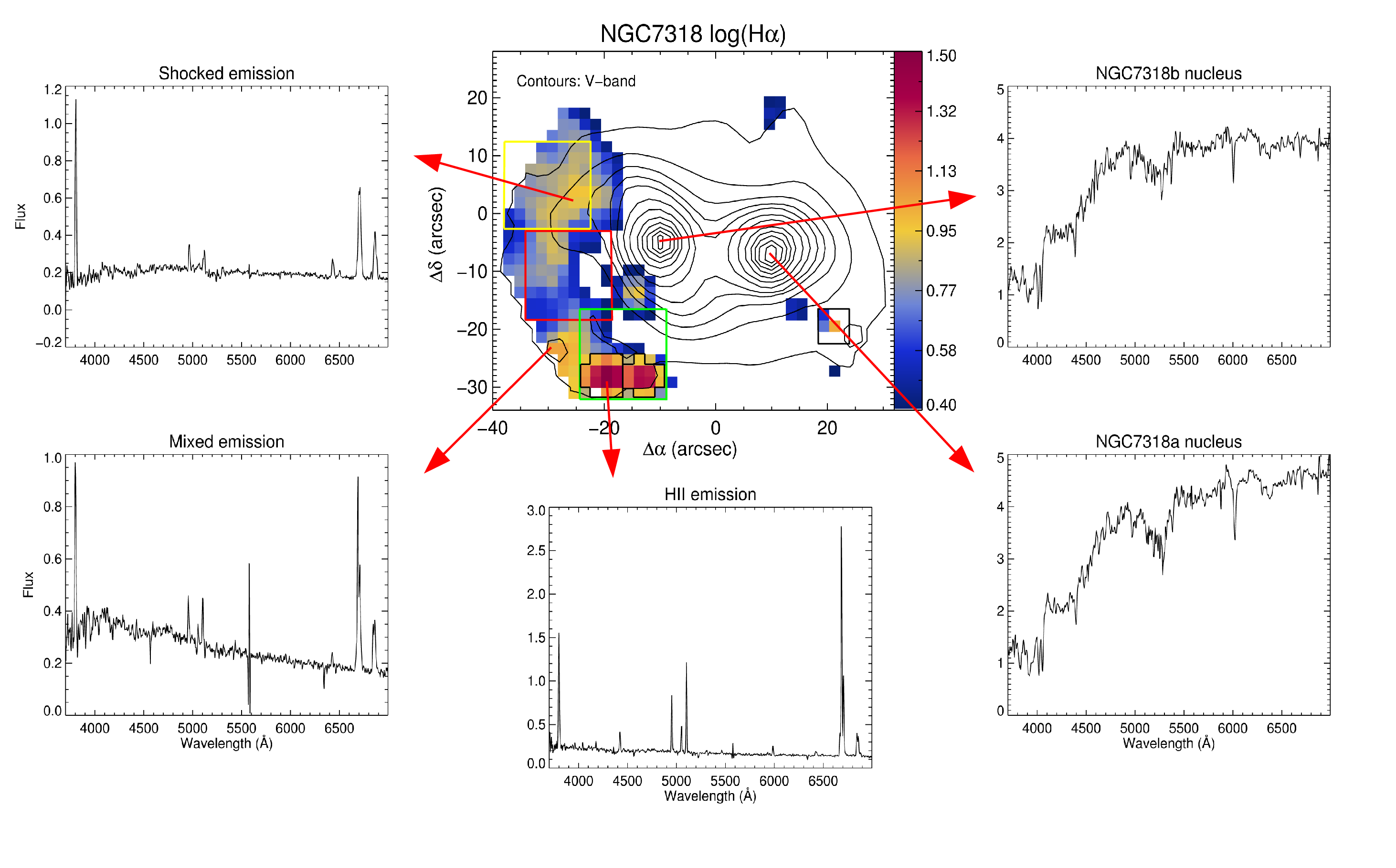}
\caption{Stacked spectra, obtained with {\sc PINGSoft}, from several key
  regions of NGC\,7318A/B. The \hh regions are delimited with black lines in the map. Yellow, red and green squares are
  \citet{2012A&A...539A.127I} pointings, overplotted for comparison. Fluxes are expressed in units of 10$^{-16}$ erg s$^{-1}$ cm$^{-2}$. (The prominent feature observed in some spectra at $\sim 5500$ \AA\ is the \ion{O}{i}~$\lambda$5577 sky emission line, the most visible at Calar Alto.) A colour version of this figure can be found in the online version of the article.}
\label{7318total}
\end{figure*}

We have overplotted on the diagnostic diagrams the predictions from the fast-shock
models with precursor photoionization of \citet{2008ApJS..178...20A}. Shock models are generated
using solar metallicity and a pre-shock density of 1 cm$^{-3}$, in agreement with
the density value derived above, and are represented for a range of
shock-velocities (v$_{s}$) and magnetic field strengths (B). Similarly we have
also overplotted on the diagnostic diagrams the predictions from the dust-free AGN
models of \citet{2004ApJS..153...75G}. AGN models are generated using metallicity double than solar, as most of the narrow-line regions in Seyfert galaxies are characterized by supersolar metallicity, with a mean value of $\sim 2Z_{\odot}$ \citep{2004ApJS..153...75G}, and an electronic
density value of 100 cm$^{-3}$, similar to the one derived from the
\sii~$\lambda\lambda$6717/6731 ratio in the nucleus. AGN models are represented for a range
of the dimensionless ionization parameter (\textit{u}) and the spectral index
($\alpha$). Both shock and AGN models were obtained using {\sc ITERA} (IDL Tool for
Emission-line Ratio Analysis; \citealt{2010NewA...15..614G}).

Regarding shock theoretical predictions, the shock plus precursor models
follow the track of the spaxels with an excellent agreement for the velocity
range of $\sim200-1000$ km s$^{-1}$. On the other hand, the observed line
ratios are inconsistent with the models that do not include the effect of the
precursor. Two of these diagrams, along with the theoretical shock models, are
presented in the top panels of Fig. \ref{7319bpts}. The wide range of shock velocities detected may indicate the presence of several kinematic components, but this fact could not be discarded nor confirmed with our data, which were not intended to perform a kinematic study of the system. With respect to the AGN
models, we first derive observational values of the ionization parameter \textit{u} from
the \oii/\oiii ratio for several regions of the galaxy, following \citet{2000MNRAS.318..462D} and obtaining a range of log~$u$ values between -2.7 and
-3.4. Taking these results into account, we observe that dust-free AGN models
fit the spaxels positions with better agreement than dusty AGN models. Two of the diagrams, with AGN models
overplotted, are presented in the bottom panels of Fig. \ref{7319bpts}.

The observed line ratios are thus consistent both with shock and AGN models,
but neither of these possibilities can explain the observational results
exclusively. Even the nucleus emission, classified as Seyfert 2, is not
plainly consistent with AGN models, which may be caused by a partial
contamination by shock emission that we observe projected on the nucleus. We
reach the conclusion that the excitation mechanism of the gas in the galaxy is
a mixture of AGN photoionization and shocks, and the resulting emission
spectra are generated by the combination of both ionization mechanisms. Thus,
while A96 scenario considered the AGN photoionization as the only responsible
mechanism for exciting the gas, the spatial resolution and wide FoV of
our data have allowed to detect the presence of shocks as another source of
ionization, and to study the coexistence of these two processes in the
observed gas emission of the galaxy.

\subsection{The galaxy interaction in NGC\,7318A/B}
\label{discussion7318}

The PPAK pointing corresponding to NGC\,7318 includes two objects: NGC\,7318A
which belongs to the group core, and NGC\,7318B which is entering the group
for the first time and colliding with it. The shock front generated by this
collision is one of the most interesting features of the SQ, but until
recently studies had been restricted to the brightest
regions. \citet{2012A&A...539A.127I}, hereafter IP12, analysed for the
first time IFS observations of the shock region in order to get a complete
mapping of physical and dynamical properties, with very interesting results
about the different kinematical components of the gas. This region has also been spectroscopically studied in detail by \citet{2014ApJ...784....1K}. Our observations, obtained with the same spectrograph but in a different configuration, have a
much wider FoV and wavelength range, although less spectral
resolution. Thus our results may confirm and complement those obtained by
IP12, giving more information by providing us with the 2D distribution of the
physical properties of the region. 

Fig. \ref{7318total} shows the H$\alpha$ emission line map with the stacked
spectra of some of the key regions of the pointing. The three pointings from
IP12 are overplotted for comparison. Gas emission is only detected in
the shock front, and the spectra of both galactic nuclei show only stellar
emission. In the south region of our pointing two \hh regions are detected,
coinciding with those mentioned by IP12 in their S pointing. Further north we
observe the shocked gas emission, with an evident strong blending in
H$\alpha$-\nii\, lines and in the \sii\, doublet. This is due to the presence of
the two different kinematical components studied by IP12, although our data do
not have enough spectral resolution to develop a similar kinematic analysis.

We extracted the stacked spectra from the two \hh regions in the SE and from a
third one in the SW. After subtracting the stellar population contribution with
{\sc FIT3D} (as described in Sec. \ref{Observations}), we measured the main
line intensities of the pure-emission spectra. Metallicities (oxygen
abundances) are estimated employing the O3N2 index:

\begin{equation}
{\rm O3N2}=\log \left(\frac{{\rm [O\,III]}\,\lambda5007/H\beta}{{\rm [N\,II]}\,\lambda6584/H\alpha}\right),
\end{equation}

\noindent
using the updated calibration by \citet{2013arXiv1307.5316M}:

\begin{equation}
12+\log\left(O/H\right) =8.53-0.22 \cdot {\rm O3N2}.
\end{equation}

Resulting values, included in Table \ref{7318table}, are lower than solar and
coincide with those obtained by IP12. Uncertainties are calculated by
propagating the errors in quadrature, taking into account the systematic
errors due to the flux calibration, and the stellar fitting. Reddening corrections were not applied here, due to the closeness of the lines involved in the ratios.

\begin{figure}
\centering
\includegraphics[bb=27 177 523 600, scale=0.43]{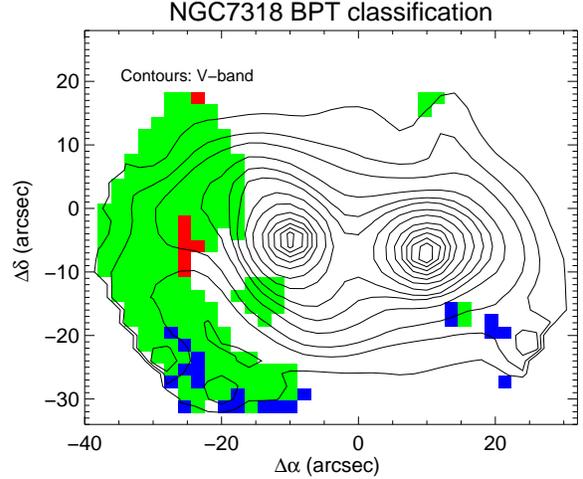}
\caption{NGC\,7318A/B spatial map with the spaxels classified according to their position in the \oiii~$\lambda$5007/H$\beta$ vs \nii~$\lambda$6584/H$\alpha$
  diagram. Spaxels classified as AGN are displayed in red, spaxels classified
  as starburst are displayed in blue, and spaxels classified as composite are
  displayed in green. A colour version of this figure can be found in the online version of the article.}
\label{7318bptmap}
\end{figure}

\begin{figure}
\centering
\includegraphics[bb=80 187 340 1030, scale=0.7]{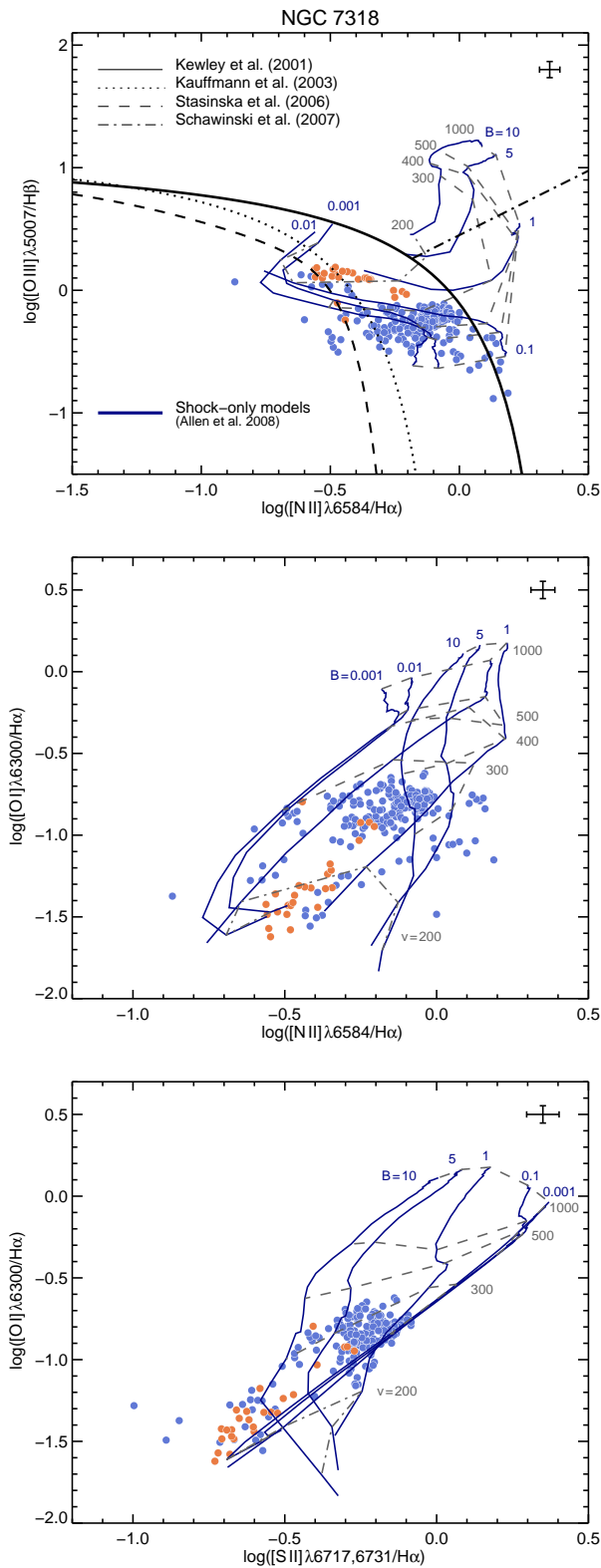}
\caption{Line ratio diagnostic diagrams, showing line ratios for independent
  spaxels in NGC\,7318A/B. Spaxels coloured in orange belong to the \hh
  regions. Overplotted as black lines in the \oiii~$\lambda$5007/H$\beta$ vs
  \nii~$\lambda$6584/H$\alpha$ diagram are empirically and theoretically
  derived separations between LINERs/Seyfert and \hh regions. Overplotted as
  coloured lines are line ratios predicted by models from \citet{2008ApJS..178...20A} for photoionization of gas by a fast shock without precursor with solar
  metallicity and low density, as described in the text. A colour version of this figure can be found in the online version of the article.}
\label{7318bpts}
\end{figure}

As in the case of NGC\,7319, diagnostic diagrams provide important information
about the shock characteristics. A map analogous to the one represented for
NGC\,7319 is shown in Fig. \ref{7318bptmap}, with the spaxels colour-coded as a function of their position in the \oiii~$\lambda$5007/H$\beta$ vs
\nii~$\lambda$6584/H$\alpha$ diagnostic diagram. Most of the spaxels are
coloured in green, as a consequence of their location in the composite zone
due to the influence of shocks in their emission. Some spaxels coloured in blue are
related to three \hh regions previously discussed. 

\begin {table}
\begin{center}
\begin{tabular}{c c}
\hline
Regions & 12\,+\,log(O/H) [O3N2] \\
\hline
1 & 8.45 $\pm$ 0.16 \\

2 & 8.39 $\pm$ 0.16 \\

3 & 8.35 $\pm$ 0.16 \\
\hline
\end{tabular}
\caption{Oxygen abundances for the \hh regions in NGC\,7318A/B pointing,
  estimated following the Marino et al. (2013) calibration based on the O3N2
  indicator.}
\label{7318table}
\end{center}
\end{table}

We have analysed the spaxels distribution in \oi~$\lambda$6300/H$\alpha$ vs \nii~$\lambda$6584/H$\alpha$ and \oi~$\lambda$6300/H$\alpha$ vs
\sii~$\lambda\lambda$6717,6731/H$\alpha$ diagnostic diagrams, that are particularly
suitable for the study of shocks, and compared our data location with the
predictions from fast-shock models by \citet{2008ApJS..178...20A}, as described in
Sec. \ref{discussion7319}. The set of models more consistent with our
observational data are those of shocks without precursor with solar
metallicity and low pre-shock density (n=0.1 cm$^{-3}$), and with velocities between 200 and 400 km s$^{-1}$. These diagrams and the overplotted models, along with the \oiii~$\lambda$5007/H$\beta$ vs
\nii~$\lambda$6584/H$\alpha$ diagram, are shown in Fig. \ref{7318bpts}. It is interesting to note that spaxels belonging to the \hh regions (coloured in orange) are located in the intermediate region between \citet{2003MNRAS.346.1055K} and \citet{2001ApJ...556..121K} demarcation lines. This fact agrees with the recent study by \citet{2013arXiv1311.7052S}, which demonstrates that {\em bona-fide} \hh regions can be found in this intermediate zone above the \citet{2003MNRAS.346.1055K} line.

IP12 analysis of diagnostic diagrams is based on their previous detection of two
kinematical components, and shows that one component overlaps
with solar metallicity shock models, while the other is shifted towards models
of lower metallicity. These results are compatible with the findings of this paper,
and discrepancies may arise from the different characteristics of the data: we
have wider FoV and wavelength range (which allows the usage of the diagrams involving the \sii~$\lambda\lambda$6717,6731 lines), but we do
not resolve kinematical components and thus we do not see the possible
existence of different trends within our data. The low metallicities obtained
for the \hh regions imply in fact that at least one of the components involved
in the collision has metallicity lower than solar. Further studies combining
gas emission lines and kinematic analysis and including information on the
metallicity are required to further analyse the characteristics of this shock
front.

\subsection{Peculiar emission in NGC\,7317}
\label{discussion7317}

The elliptical galaxy NGC\,7317 shows no significant gas emission by
itself. Nevertheless, we do detect a peculiar emission-line source within this pointing:
a region located at the southwest of NGC\,7317 that can also be observed in
the HST image (see Fig. \ref{pointings}), and at first sight may be
confused as a background galaxy. An analysis of the spectra of this region reveals
that its redshift is concordant with that of NGC\,7317, therefore this
emitting region is located close to the elliptical galaxy. The maximum visible length of the region is of $\sim$\,8 arcsec, which, adopting a distance of 93.3 Mpc to NGC\,7317 (taken from NED and
confirmed spectroscopically), corresponds to a physical length of $\sim$\,0.35 Mpc. 

\begin{figure}
\centering
\includegraphics[bb=46 188 536 580, scale=0.45]{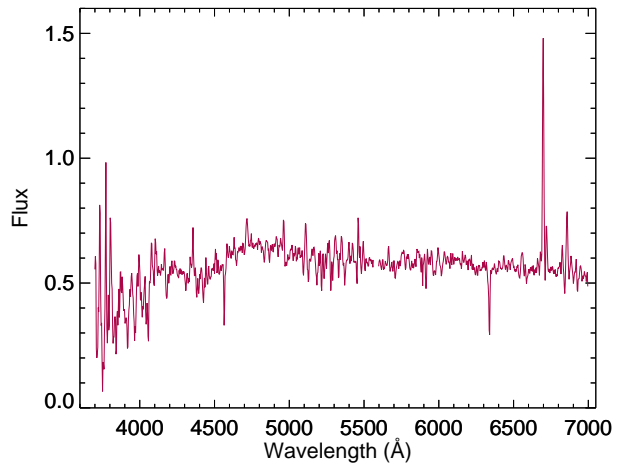}
\caption{Integrated emission spectrum from the emitting region detected southwest of NGC\,7317. Flux is expressed in units of 10$^{-16}$ erg s$^{-1}$ cm$^{-2}$.}
\label{7317hiispectrum}
\end{figure}

\begin{figure*}
\centering
\includegraphics[bb=-15 0 865 355, scale=0.60]{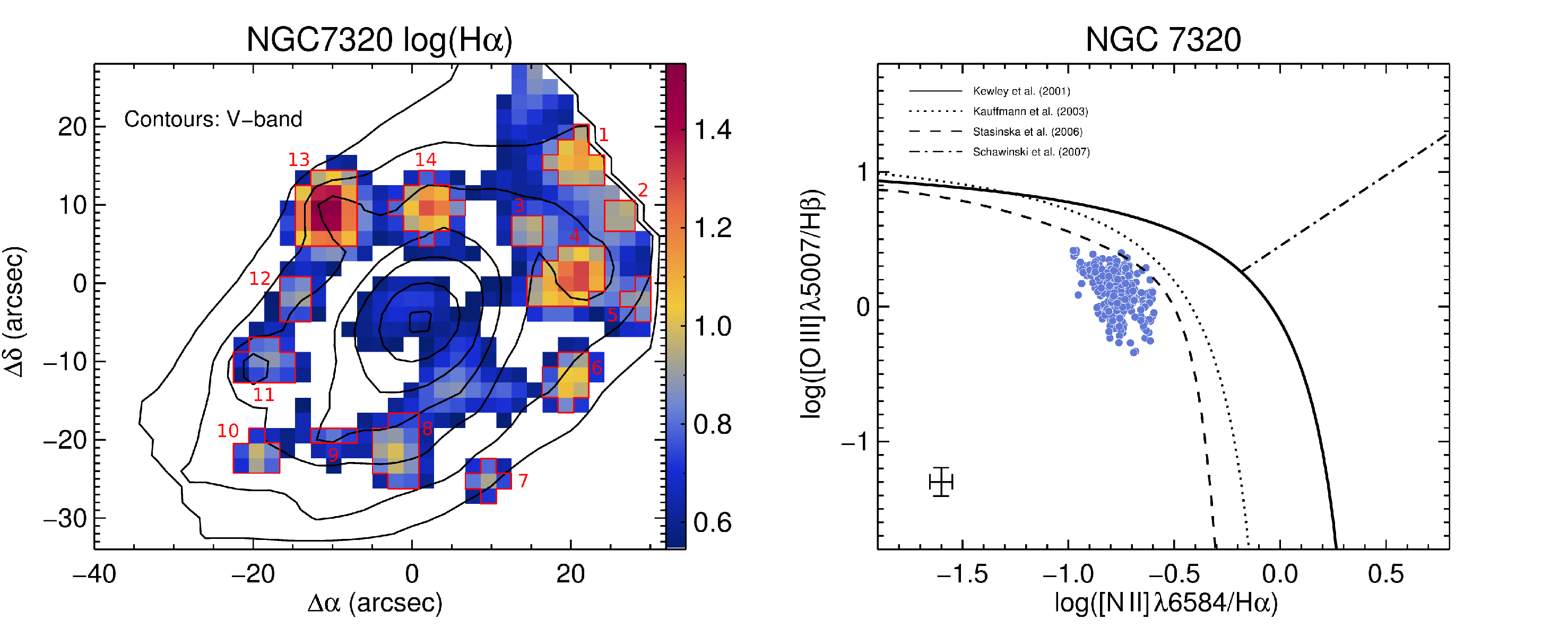}
\caption{
  Left: H$\alpha$ emission line map of NGC\,7320, with the extracted \hh regions indicated with red contours. Right: \oiii~$\lambda$5007/H$\beta$ vs \nii~$\lambda$6584/H$\alpha$ diagram, showing line ratios for independent spaxels in NGC\,7320. A colour version of this figure can be found in the online version of the article.}
\label{7320figures}
\end{figure*}

We extracted the integrated spectrum from this region, showed in Fig.\ref{7317hiispectrum}, and after subtracting the stellar continuum (see Sec. \ref{Observations}), we analysed the residual gas
emission spectrum. Metallicity was estimated following the procedure explained in Sec. \ref{discussion7318},
obtaining a value of 8.42\,$\pm$\,0.03, which is quantitatively very close to
the abundance derived for the \hh regions in NGC\,7318.

We obtained the H$\alpha$ luminosity from the reddening corrected H$\alpha$
flux, considering the mentioned distance to NGC\,7317. The H$\alpha$ flux was corrected for reddening using the Balmer decrement according to the reddening function of \citet{1989ApJ...345..245C}, assuming $R \equiv A_V /E(B - V ) = 3.1$. Theoretical value for the intrinsic line ratio H$\alpha$/H$\beta$ was
taken from \citet{2006agna.book.....O}, assuming case B recombination
(optically thick in all the Lyman lines), an electron density of $n_{e}$=100
cm$^{-3}$ and an electron temperature $T_{e}$=10$^{4}$K. The logarithmic
reddening coefficient, c(H$\beta$), was estimated using the H$\alpha$ and
H$\beta$ lines in each spectrum. The relation between the H$\alpha$ luminosity and the number of ionizing Lyman continuum photons given by \citet{1995ApJ...439..604G} gives the total number of ionizing photons. Using the total number of ionizing photons per unit mass provided by the PopStar models \citep{2009MNRAS.398..451M} for a Zero Age Main Sequence (ZAMS) with Salpeter initial mass
function (IMF) with lower and upper mass limits of 1 and 100 M$_{\odot}$ and
$Z=0.008$, we estimate an ionizing cluster mass of $2.2 \times 10^4
M_{\odot}$. This is a lower limit of the ionizing mass, as we are considering
an unevolved stellar population with no photon escape and no dust absorption. 

The present work is focused on the ionized gas emission, but as a first approach to this non previous studied region we have applied the Starlight synthesis spectral code \citep{2005MNRAS.358..363C} to its extracted spectrum in order to characterize its stellar population. We have obtained that about 80\% (in mass) of the stars has an age of $10^{10}$ years. This old population indicates that the region should had been formed inside one of the galaxies of the group, and that it was separated from the galaxy afterwards, maybe during some of the interacting episodes within the group.

The nature of this star cluster is intriguing, but may argue that its origin is related
to the interacting history of the group, and that it might be a tidal debris tail or the product of a
tidal instability which is now orbiting NGC\,7317.

\subsection{The foreground galaxy NGC\,7320}
\label{discussion7320}

This galaxy lost its membership to the Stephan's Quintet when it was found to
be a foreground object \citep{1961ApJ...134..244B}, but its presence in the
data frames is unavoidable. Therefore it is necessary to know its
characteristics, in order to disentangle its emission from that coming from
the real members of the group.

The galactic structure is clearly reproduced in the emission-line ratio
maps. Fig. \ref{7320figures} shows the H$\alpha$ emission map, that traces the
distribution of the star-forming regions in the galaxy and displays its spiral
structure, with multiple \hh regions and \hh complexes of different sizes and
morphology. The presence of diffuse emission between the \hh regions is also
noticeable. In the \nii~$\lambda$6583/H$\alpha$ and
\oiii~$\lambda\lambda$4959,5007/H$\beta$ maps, displayed in Fig. \ref{maps}, it can
be observed that the \nii~$\lambda$6583/H$\alpha$ ratio has lower values in
the central part of the \hh regions, while \oiii~$\lambda\lambda$4959,5007/H$\beta$
ratio shows an opposite behaviour. Given that the \nii~$\lambda$6583 emission
originates in singly ionized regions, between the fully ionized and the
partially ionized zones, the \nii~$\lambda$6583/H$\alpha$ ratio traces the
changes in the local ionization, while the \oiii~$\lambda\lambda$4959,5007 originates
in the fully ionized zones, tracing the strength of the ionization. Therefore,
this distribution may indicate that the ionization is stronger in the outer
parts of the galaxy structure.

The diagnostic diagrams show the expected spaxels distribution. The
\oiii~$\lambda$5007/H$\beta$ vs \nii~$\lambda$6584/H$\alpha$ BPT diagram (included
in Fig. \ref{7320figures}), as an example, shows that all the spaxels are located
in the starburst region, even considering the most restrictive parametrization
\citep{2006MNRAS.371..972S}. Other diagrams show the same tendency.

The stacked spectra from all the \hh regions (indicated in Fig. \ref{7320figures}) were extracted, and after
accomplishing the fitting process with {\sc FIT3D} (see
Sec. \ref{Observations}) we obtained the measurements of the main line fluxes
from the residual spectra. The observed intensities are normalized to the flux
of H$\beta$ in units of 10$^{-16}$ erg s$^{-1}$ cm$^{-2}$ \AA$^{-1}$, and are
shown in Tables \ref{7320table1} to \ref{7320table3} in the columns labelled as
F($\lambda$)/H$\beta$. Thereafter, the observed line intensities were
corrected for reddening as explained in Sec.\ref{discussion7317}. Tables \ref{7320table1} to \ref{7320table3}
show the reddening corrected emission line fluxes for each integrated
spectrum, designated by the I($\lambda$)/I(H$\beta$) columns. Uncertainties
are calculated as mentioned in Sec. \ref{discussion7318}.

\begin{figure}
\centering
\includegraphics[bb=46 189 540 600, scale=0.45]{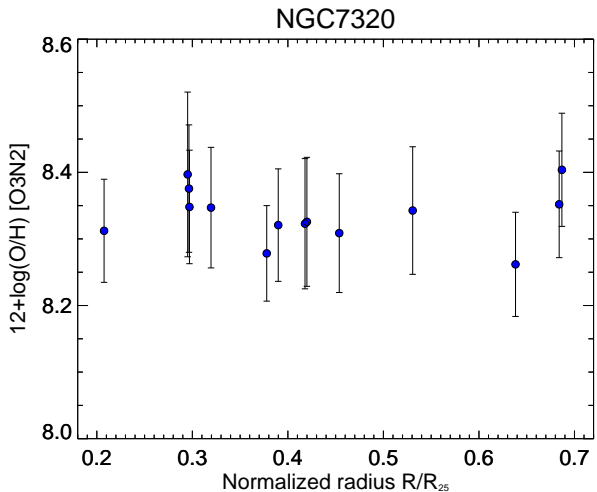}
\caption{
  Gas-phase oxygen abundance obtained for the \hh regions in NGC\,7320 as a
  function of the deprojected galactocentric distance normalized to the optical radius.
  \label{7320metvsdist}}
\end{figure}

We estimate the \hh regions metallicities from the line intensities and their
uncertainties as explained in Sec. \ref{discussion7318}. Obtained values are
included in Table\ref{7320mets}. We plot the metallicity of the \hh
regions as a function of their deprojected distances to the nucleus normalized
to the optical radius (R$_{25}$; \citealt{1991S&T....82Q.621D}) in
Fig. \ref{7320metvsdist}. A very subtle metallicity gradient is observed between the
distances of 0.3 and 0.65 r/R$_{25}$, while the internal region of the galaxy
presents a lower metallicity value, both phenomena described in \citet{2012A&A...538A...8S}. Two puzzling regions show high metallicity values although being
located at high galactocentric distances, that might suggest an average flat
oxygen abundance for the entire galaxy disk.

The mean value of the metallicities from the \hh regions allows us to estimate
the value of the stellar mass, using the mass-metallicity relation from
\citet{2013A&A...554A..58S} based on CALIFA data \citep{2012A&A...538A...8S}. We find that the mass galaxy is of the order of 15 $\times$ 10${^8}$ M$_{\odot}$, which indicates that it is a low mass galaxy.


\section{Summary and conclusions}
\label{conclusions}

We have analysed IFS observations of the Stephan's Quintet provided by the
PINGS survey. The data were composed by four pointings, each one with a FoV of
74x65 arcsec$^2$ and containing 331 optical fibres that provided spectra with a
wavelength range between 3700 and 7100 \AA\ and with a spectral resolution of
10\,\AA\ FWHM. The pointings were centred on the bulges of NGC\,7317, NGC\,7319,
NGC\,7320 and in configuration over NGC\,7318A and NGC\,7318B. The stellar population of each spaxel spectrum was subtracted
by the application of the {\sc FIT3D} software, in order to obtain the pure
emission spectra. Thus we were able to analyse the gas emission from all the
galaxies with an unprecedented spatial resolution.
For three of the pointings (except NGC\,7317) we obtained maps of the main
emission-line ratios, and represented diagnostic diagrams where the spatial
resolution of the data allowed to locate separately each one of the spaxels
from the datacube. As a consequence we could detect and study different
ionization sources coexisting in the same region.

In NGC\,7319, the analysis was focused in the active nucleus and its
surroundings. The stacked spectrum from the nucleus confirmed its
classification as Seyfert 2. \citet{1996AJ....111..140A} already reported the
presence of a high-velocity outflow within this galaxy, based on the detection
of a region in the south-west of the nucleus where the gas is blueshifted from
the systemic velocity by 300 km s$^{-1}$ in average and 500 km s$^{-1}$ at
maximum. Our data have not enough spectral resolution to develop a kinematic
analysis, but we have studied the effect of this outflow in the measured
emission-line ratios. We have compared the spaxels distribution in the
diagnostic diagrams with the predictions from fast-shock models by \citet{2008ApJS..178...20A} and from AGN models by \citet{2004ApJS..153...75G}, obtaining that the sum
of both ionization sources is necessary to explain the observed ratios. While
\citet{1996AJ....111..140A} considered that photoionization by AGN was the only
responsible mechanism of the excitation of the gas, the wide FoV and the
spatial resolution of our results reveal the presence of shocks with precursor
with a velocity range of $\sim$200-1000 km s$^{-1}$ coexisting with the AGN
photoionization.

In the case of the NGC\,7318A/B pointing, gas line emission arises almost
exclusively from the shock front. Shocked gas
emission spectra confirmed the presence of several kinematic components, as
reported in \citet{2012A&A...539A.127I}. Two \hh regions were detected in
the south, and their gas oxygen abundances were estimated using the updated O3N2 calibrator by \citet{2013arXiv1307.5316M}, corresponding to a subsolar metallicity. The comparison of the spaxels
distribution in the diagnostic diagrams with the predictions from fast-shock
models by \citet{2008ApJS..178...20A} suggests that our observational data are
consistent with shocks without precursor with solar metallicity and low
pre-shock density, in a range of velocities of $\sim200-400$ km s$^{-1}$. Our
results confirm and complement those obtained by \citet{2012A&A...539A.127I}, specially in terms of the subsolar metallicities obtained for the \hh regions. 

NGC\,7317 showed no significant gas emission by itself, and we could not
obtain maps or diagnostic diagrams for this pointing. Nevertheless, an
emitting region was detected southwest to the galaxy. Its integrated spectrum
was extracted, and we estimated its metallicity and the ionizing star cluster
mass. We postulate that this cluster might be the result of a tidal
interaction with the rest of the members in the SQ group.

Finally, as a by-product of the IFS observations, we analyse the 2D distribution of the star-forming regions in the foreground galaxy NGC\,7320. The spectra from
all the \hh regions were extracted, and the metallicities were calculated from
the reddening corrected line intensities. The representation of these
metallicity values as a function of the deprojected distances to the nucleus
shows a weak metallicity gradient along r/R$_{25}$ with a metallicity fall in the
internal region, as expected \citep{2013A&A...554A..58S}, and two puzzling regions with high metallicity
values at high galactocentric distances. Using the mass-metallicity relation
from \citet{2013A&A...554A..58S} we estimate that NGC\,7320 is a low mass galaxy, with
a mass of the order of 15 $\times$ 10${^8}$ M$_{\odot}$.


\section*{Acknowledgements}

We thank the anonymous referee for all the comments and suggestions that improved the content of the paper. We would like to thank Rub\'{e}n Garc\'{i}a-Benito for his support and useful
discussions during the preparation of this manuscript.
We acknowledge financial support for the ESTALLIDOS collaboration by the
Spanish Ministerio de Ciencia e Innovaci\'{o}n under grant AYA2010-21887-C04-03.
We acknowledge financial support from the Marie Curie FP7-PEOPLE-2013-IRSES scheme, under the SELGIFS collaboration (Study of Emission-Line Galaxies with Integral-Field Spectroscopy). M.~R.~B. acknowledges financial support by the Spanish Ministerio de Econom\'{i}a y Competitividad under the FPI fellowships programme. 
F.~F.~R.~O. acknowledges the Mexican National Council for Science and
Technology (CONACYT) for financial support under the programme Estancias
Posdoctorales y Sab\'{a}ticas al Extranjero para la Consolidaci\'{o}n de
Grupos de Investigaci\'{o}n, 2010-2012.


\bibliographystyle{mn2e}



\begin{table*}
\begin{center}  

\begin{tabular}{l c c c c c c c c }
\hline
& \multicolumn{2}{c}{1} &\multicolumn{2}{c}{2} &\multicolumn{2}{c}{3} &\multicolumn{2}{c}{4} \\
& F($\lambda$)/H$\beta$ & I($\lambda$)/H$\beta$ & F($\lambda$)/H$\beta$ & I($\lambda$)/H$\beta$ & F($\lambda$)/H$\beta$ & I($\lambda$)/H$\beta$ & F($\lambda$)/H$\beta$ & I($\lambda$)/H$\beta$ \\ 
\hline

\oii~$\lambda$3727 & 4.57$\pm$0.97 & 4.57$\pm$0.97 & 4.57$\pm$0.97 & 4.84$\pm$1.21 & 3.08$\pm$0.67 & 3.26$\pm$0.83 & 3.14$\pm$0.65 & 3.42$\pm$0.83\\

\oiii~$\lambda$4959 & 0.53$\pm$0.09 & 0.53$\pm$0.09 & 0.30$\pm$0.06 & 0.30$\pm$0.06 & 0.21$\pm$0.05 & 0.21$\pm$0.05 & 0.49$\pm$0.08 & 0.49$\pm$0.08 \\

\oiii~$\lambda$5007 & 1.56$\pm$0.23 & 1.56$\pm$0.23 & 1.13$\pm$0.17 & 1.12$\pm$0.17 & 0.91$\pm$0.14 & 0.90$\pm$0.14 & 1.50$\pm$0.21 & 1.49$\pm$0.21 \\

\hei~$\lambda$5876 & 0.06$\pm$0.02 & 0.06$\pm$0.02 & -- & -- & -- & -- & -- & -- \\

\oi~$\lambda$6300 & 0.09$\pm$0.02 & 0.09$\pm$0.02 & 0.11$\pm$0.02 & 0.11$\pm$0.02 & 0.06$\pm$0.02 & 0.06$\pm$0.02 & 0.05$\pm$0.01 & 0.05$\pm$0.01 \\

\nii~$\lambda$6548 & 0.16$\pm$0.05 & 0.16$\pm$0.05 & 0.15$\pm$0.05 & 0.14$\pm$0.05 & 0.17$\pm$0.05 & 0.16$\pm$0.05 & 0.14$\pm$0.04 & 0.13$\pm$0.04 \\

H$\alpha$ $\lambda$6562 & 2.82$\pm$0.36 & 2.82$\pm$0.36 & 3.03$\pm$0.37 & 2.87$\pm$0.50 & 3.03$\pm$0.38 & 2.87$\pm$0.51 & 3.11$\pm$0.37 & 2.87$\pm$0.48 \\

\nii~$\lambda$6583 & 0.46$\pm$0.08 & 0.46$\pm$0.08 & 0.44$\pm$0.07 & 0.42$\pm$0.09 & 0.50$\pm$0.08 & 0.48$\pm$0.10 & 0.42$\pm$0.06 & 0.39$\pm$0.08 \\

\sii~$\lambda$6717 & 0.56$\pm$0.09 & 0.56$\pm$0.09 & 0.58$\pm$0.09 & 0.55$\pm$0.11 & 0.53$\pm$0.08 & 0.50$\pm$0.10 & 0.48$\pm$0.07 & 0.44$\pm$0.08 \\

\sii~$\lambda$6731 & 0.38$\pm$0.07 & 0.38$\pm$0.07 & 0.48$\pm$0.08 & 0.45$\pm$0.09 & 0.42$\pm$0.07 & 0.40$\pm$0.08 & 0.39$\pm$0.06 & 0.35$\pm$0.07 \\

& & & & & & & & \\
F(H$\beta$) $\lambda$4861 & 44.97 & & 13.82 & & 13.18 & & 80.44 & \\

\hline
\end{tabular}  
\caption{Integrated line intensities for NGC\,7320, regions 1 to 4. The first column correspond to the emission line identification, with the rest-frame wavelength. The F($\lambda$)/H$\beta$ column corresponds to the observed flux, while the I($\lambda$)/H$\beta$ to the reddening corrected values; normalised to H$\beta$. The observed fluxes in H$\beta$ are expressed in units of 10$^{-16}$ erg s$^{-1}$ cm$^{-2}$.}
\label{7320table1}
\end{center}
\end{table*}


\begin{table*}
\begin{center}

\begin{tabular}{l c c c c c c c c }
\hline
& \multicolumn{2}{c}{5} &\multicolumn{2}{c}{6} &\multicolumn{2}{c}{7} &\multicolumn{2}{c}{8} \\
& F($\lambda$)/H$\beta$ & I($\lambda$)/H$\beta$ & F($\lambda$)/H$\beta$ & I($\lambda$)/H$\beta$ & F($\lambda$)/H$\beta$ & I($\lambda$)/H$\beta$ & F($\lambda$)/H$\beta$ & I($\lambda$)/H$\beta$ \\ 
\hline

\oii~$\lambda$3727 & 4.27$\pm$0.90 & 4.64$\pm$1.14 & 4.07$\pm$0.77 & 4.23$\pm$0.94 & 4.77$\pm$0.92 & 4.77$\pm$0.92 & 4.27$\pm$0.85 & 4.57$\pm$1.09 \\

\oiii~$\lambda$4959 & 0.19$\pm$0.04 & 0.19$\pm$0.04 & 0.69$\pm$0.10 & 0.69$\pm$0.10 & 0.31$\pm$0.05 & 0.31$\pm$0.05 & 0.48$\pm$0.08 & 0.48$\pm$0.08 \\

\oiii~$\lambda$5007 & 0.68$\pm$0.10 & 0.68$\pm$0.10 & 2.08$\pm$0.27 & 2.07$\pm$0.27 & 0.95$\pm$0.13 & 0.95$\pm$0.13 & 1.42$\pm$0.20 & 1.41$\pm$0.20 \\

\hei~$\lambda$5876 & -- & -- & -- & -- & -- & -- & -- & -- \\

\oi~$\lambda$6300 & -- & -- & 0.06$\pm$0.02 & 0.05$\pm$0.02 & 0.08$\pm$0.02 & 0.08$\pm$0.02 & -- & -- \\

\nii~$\lambda$6548 & 0.18$\pm$0.04 & 0.17$\pm$0.04 & 0.12$\pm$0.02 & 0.11$\pm$0.02 & 0.13$\pm$0.02 & 0.13$\pm$0.02 & 0.16$\pm$0.05 & 0.15$\pm$0.05 \\

H$\alpha$ $\lambda$6562 & 3.10$\pm$0.36 & 2.87$\pm$0.47 & 2.97$\pm$0.33 & 2.87$\pm$0.44 & 2.83$\pm$0.32 & 2.83$\pm$0.32 & 3.06$\pm$0.37 & 2.87$\pm$0.49 \\

\nii~$\lambda$6583 & 0.53$\pm$0.07 & 0.49$\pm$0.09 & 0.33$\pm$0.04 & 0.32$\pm$0.05 & 0.38$\pm$0.05 & 0.38$\pm$0.05 & 0.47$\pm$0.08 & 0.44$\pm$0.09 \\

\sii~$\lambda$6717 & 0.68$\pm$0.09 & 0.63$\pm$0.11 & 0.36$\pm$0.04 & 0.34$\pm$0.06 & 0.50$\pm$0.06 & 0.50$\pm$0.06 & 0.91$\pm$0.13 & 0.85$\pm$0.16 \\

\sii~$\lambda$6731 & 0.52$\pm$0.07 & 0.47$\pm$0.09 & 0.27$\pm$0.04 & 0.26$\pm$0.05 & 0.38$\pm$0.05 & 0.38$\pm$0.05 & 0.46$\pm$0.08 & 0.43$\pm$0.09 \\

& & & & & & & & \\
F(H$\beta$) $\lambda$4861 & 11.47 & & 31.69 & & 13.72 & & 37.58 & \\

\hline

\end{tabular}
\caption{Integrated line intensities for NGC\,7320, regions 5 to 8. The first column correspond to the emission line identification, with the rest-frame wavelength. The F($\lambda$)/H$\beta$ column corresponds to the observed flux, while the I($\lambda$)/H$\beta$ to the reddening corrected values; normalised to H$\beta$. The observed fluxes in H$\beta$ are expressed in units of 10$^{-16}$ erg s$^{-1}$ cm$^{-2}$.}
\label{7320table2}
\end{center}
\end{table*}


\begin{table*}

\begin{center}

\begin{tabular}{l c c c c c c }
\hline
& \multicolumn{2}{c}{9} &\multicolumn{2}{c}{10} &\multicolumn{2}{c}{11}\\
& F($\lambda$)/H$\beta$ & I($\lambda$)/H$\beta$ & F($\lambda$)/H$\beta$ & I($\lambda$)/H$\beta$ & F($\lambda$)/H$\beta$ & I($\lambda$)/H$\beta$ \\ 
\hline

\oii~$\lambda$3727 & 3.89$\pm$0.91 & 3.89$\pm$0.91 & 4.67$\pm$0.97 & 5.40$\pm$1.32 & 6.28$\pm$1.28 & 6.75$\pm$1.67 \\

\oiii~$\lambda$4959 & 0.32$\pm$0.08 & 0.32$\pm$0.08 & 0.45$\pm$0.08 & 0.44$\pm$0.08 & 0.45$\pm$0.09 & 0.45$\pm$0.09 \\

\oiii~$\lambda$5007 & 0.71$\pm$0.14 & 0.71$\pm$0.14 & 1.43$\pm$0.21 & 1.41$\pm$0.21 & 1.34$\pm$0.22 & 1.33$\pm$0.22 \\

\hei~$\lambda$5876 & 0.32$\pm$0.06 & 0.32$\pm$0.06 & -- & -- & 0.41$\pm$0.10 & 0.39$\pm$0.10 \\

\oi~$\lambda$6300 & -- & -- & -- & -- & 0.16$\pm$0.03 & 0.15$\pm$0.03 \\

\nii~$\lambda$6548 & 0.14$\pm$0.06 & 0.14$\pm$0.06 & 0.17$\pm$0.03 & 0.14$\pm$0.03 & 0.19$\pm$0.04 & 0.18$\pm$0.05 \\

H$\alpha$ $\lambda$6562 & 2.52$\pm$0.36 & 2.52$\pm$0.36 & 3.29$\pm$0.40 & 2.87$\pm$0.49 & 3.07$\pm$0.41 & 2.87$\pm$0.54 \\

\nii~$\lambda$6583 & 0.41$\pm$0.09 & 0.41$\pm$0.09 & 0.48$\pm$0.07 & 0.42$\pm$0.07 & 0.56$\pm$0.08 & 0.52$\pm$0.10 \\

\sii~$\lambda$6717 & 0.68$\pm$0.12 & 0.68$\pm$0.12 & 0.62$\pm$0.08 & 0.54$\pm$0.10 & 0.82$\pm$0.12 & 0.76$\pm$0.15 \\

\sii~$\lambda$6731 & 0.40$\pm$0.08 & 0.40$\pm$0.08 & 0.41$\pm$0.06 & 0.36$\pm$0.07 & 0.59$\pm$0.09 & 0.55$\pm$0.11 \\

& & & & & &\\
F(H$\beta$) $\lambda$4861 & 11.64 & & 17.34 & & 26.22 & \\

\hline

\end{tabular} 
\caption{Integrated line intensities for NGC\,7320, regions 9 to 11. The first column correspond to the emission line identification, with the rest-frame wavelength. The F($\lambda$)/H$\beta$ column corresponds to the observed flux, while the I($\lambda$)/H$\beta$ to the reddening corrected values; normalised to H$\beta$. The observed fluxes in H$\beta$ are expressed in units of 10$^{-16}$ erg s$^{-1}$ cm$^{-2}$.}
\label{7320table3}

\end{center}

\end{table*}


\begin{table*}

\begin{center}

\begin{tabular}{l c c c c c c }
\hline
& \multicolumn{2}{c}{12} &\multicolumn{2}{c}{13} &\multicolumn{2}{c}{14}\\
& F($\lambda$)/H$\beta$ & I($\lambda$)/H$\beta$ & F($\lambda$)/H$\beta$ & I($\lambda$)/H$\beta$ & F($\lambda$)/H$\beta$ & I($\lambda$)/H$\beta$ \\ 
\hline

\oii~$\lambda$3727 & 3.79$\pm$0.79 & 4.37$\pm$1.08 & 3.66$\pm$0.67 & 3.66$\pm$0.67 & 3.84$\pm$0.73 & 4.17$\pm$0.93 \\

\oiii~$\lambda$4959 & 0.44$\pm$0.08 & 0.43$\pm$0.08 & 0.64$\pm$0.08 & 0.64$\pm$0.08 & 0.52$\pm$0.08 & 0.52$\pm$0.08 \\

\oiii~$\lambda$5007 & 1.34$\pm$0.20 & 1.32$\pm$0.20 & 2.01$\pm$0.25 & 2.01$\pm$0.25 & 1.76$\pm$0.23 & 1.74$\pm$0.23 \\

\hei~$\lambda$5876 & -- & -- & 0.02$\pm$0.01 & 0.02$\pm$0.01 & -- & - \\

\oi~$\lambda$6300 & 0.06$\pm$0.02 & 0.06$\pm$0.02 & 0.04$\pm$0.01 & 0.04$\pm$0.01 & 0.07$\pm$0.01 & 0.07$\pm$0.01 \\

\nii~$\lambda$6548 & 0.21$\pm$0.04 & 0.18$\pm$0.04 & 0.12$\pm$0.02 & 0.12$\pm$0.02 & 0.16$\pm$0.03 & 0.16$\pm$0.03 \\

H$\alpha$ $\lambda$6562 & 3.28$\pm$0.41 & 2.87$\pm$0.51 & 2.77$\pm$0.28 & 2.77$\pm$0.28 & 3.10$\pm$0.35 & 2.87$\pm$0.45 \\

\nii~$\lambda$6583 & 0.60$\pm$0.08 & 0.52$\pm$0.10 & 0.36$\pm$0.04 & 0.36$\pm$0.04 & 0.51$\pm$0.06 & 0.47$\pm$0.08 \\

\sii~$\lambda$6717 & 0.72$\pm$0.10 & 0.62$\pm$0.12 & 0.40$\pm$0.04 & 0.40$\pm$0.04 & 0.63$\pm$0.08 & 0.58$\pm$0.10 \\

\sii~$\lambda$6731 & 0.49$\pm$0.07 & 0.42$\pm$0.08 & 0.27$\pm$0.03 & 0.27$\pm$0.03 & 0.46$\pm$0.06 & 0.42$\pm$0.07 \\

& & & & & &\\
F(H$\beta$) $\lambda$4861 & 15.11 & & 133.6 & & 54.69 &  \\

\hline

\end{tabular} 
\caption{Integrated line intensities for NGC\,7320, regions 12 to 14. The first column correspond to the emission line identification, with the rest-frame wavelength. The F($\lambda$)/H$\beta$ column corresponds to the observed flux, while the I($\lambda$)/H$\beta$ to the reddening corrected values; normalised to H$\beta$. The observed fluxes in H$\beta$ are expressed in units of 10$^{-16}$ erg s$^{-1}$ cm$^{-2}$.}
\label{7320table4}

\end{center}

\end{table*}

\begin{table*}
\begin{center}

\begin{tabular}{c c c c c c c c}
\hline
& 1 & 2 & 3 & 4 & 5 & 6 & 7 \\
\hline
12\,+\,log(O/H) & 8.32 $\pm$ 0.09 & 8.34 $\pm$ 0.09 & 8.38 $\pm$ 0.10 & 8.31 $\pm$ 0.09 & 8.40 $\pm$ 0.08 & 8.26 $\pm$ 0.08 & 8.35 $\pm$ 0.08 \\

\hline
& 8 & 9 & 10 & 11 & 12 & 13 & 14 \\
\hline
12\,+\,log(O/H) & 8.33 $\pm$ 0.10 & 8.40 $\pm$ 0.12 & 8.32 $\pm$ 0.08 & 8.35 $\pm$ 0.09 & 8.35 $\pm$ 0.09 & 8.28 $\pm$ 0.07 & 8.31 $\pm$ 0.08 \\
\hline
\end{tabular}
\caption{Oxygen abundances for the \hh regions in NGC\,7320 pointing,
  estimated following the Marino et al. (2013) calibration, using O3N2 indicator.}
\label{7320mets}

\end{center}

\end{table*}

\label{lastpage}

\end{document}